\documentclass[a4paper,11pt]{article}
\pdfoutput=1 

\usepackage{jheppub} 

\usepackage[T1]{fontenc} 

\usepackage{subcaption}

\title{\boldmath Sensitivity of reactor experiments to nonstandard neutrino interactions in beta decay rates}

\author[]{Andrew D. Santos}

\affiliation[]{Department of Physics \& Astronomy, Purdue University, West Lafayette, IN 47906, USA}

\emailAdd{santos30@purdue.edu}

\abstract{We frame beta-minus decay rate perturbations in the context of charged-current (CC) nonstandard neutrino interactions (NSI). In particular, we first outline one NSI parameterization for modeling the CC NSI. Then, we demonstrate that the strength of the NSI constrained by beta-minus decay data is comparable to previously reported bounds on general CC NSI at $\mathcal{O}(10^{-4})$ to $\mathcal{O}(10^{-2})$. After discussing possible parameters involved in beta-minus decay NSI, we establish a working framework to probe potentially new physics in nuclear decay rates. Finally, we determine that current nuclear reactor technology could be used for experiments that are sensitive to these NSI parameters. These would include NSI contributions from two types of parameters: (i) the relative NSI effects from the three neutrino flavors and (ii) the change in flux of electron neutrinos through a decaying sample.}

\keywords{Neutrino Physics, Beyond Standard Model}

\begin{document} 
\maketitle
\flushbottom

\section{Introduction}

The Standard Model has been one of the most powerful tools for studying particle physics. Yet, there remain many open questions about phenomena that seemingly cannot be explained within its conventional framework. Such physics is beyond the Standard Model (i.e., BSM physics). For example, the discovery of neutrino oscillations \cite{super-kamiokande_collaboration_evidence_1998} arising from mismatching flavor and mass eigenstates highlighted the need for BSM searches---or at least for revising the Standard Model to accommodate for new physics. As research has concentrated on this area, one expanding body of work has included studies of ``nonstandard neutrino interactions" (NSI). Operating within this framework has been a popular way of approaching BSM neutrino physics.

Toward the beginning of the 2000's, an important study of general NSI bounds came with Ref.\ \cite{biggio_general_2009} along with an updated consideration of these authors' loop bounds with Ref.\ \cite{biggio_loop_2009}. Charged-current (CC) NSI in these studies were constrained below $\mathcal{O}(10^{-1})$ using measurements of the Fermi constant $G_F$ and of the Cabibbo-Kobayashi-Maskawa (CKM) quark matrix. Another important study for CC NSI a few years later looked at low-energy and collider experiment bounds \cite{cirigliano_non-standard_2013}. A summary \cite{ohlsson_status_2013} of phenomenological bounds---including a review of NSI---reported direct bounds on matter NSI, direct bounds on production and detection NSI, bounds on NSI in neutrino cross-sections, and bounds on NSI using accelerators. One of the most recent status reports on NSI analyzed bounds from several more sources \cite{dev_neutrino_2019}. Indeed, this research has expanded to touch much of neutrino physics and astrophysics.

This NSI framework has been applied to neutrino experiments at all stages of development and operation. Ref.\ \cite{liao_nonstandard_2017} studied NSI at the future Hyper-Kamiokande experiment and Jiangmen Underground Neutrino Observatory (JUNO). NSI at the Deep Underground Neutrino Experiment (DUNE), the Tokai-to-Hyper-Kamiokande (T2HK) experiment, and the Tokai-to-Hyper-Kamiokande-and-Korea (T2HKK) experiment were studied in Ref.\ \cite{liao_nonstandard_2017-1}. Refs.\ \cite{de_gouvea_non-standard_2016} and \cite{bakhti_sensitivities_2017} focused solely on NSI at DUNE. Finally, Ref.\ \cite{papoulias_recent_2019} considered the effect of recent and future experiments on both standard and nonstandard physics in nuclei.

The NSI can be manifested in a diverse set of observable quantities in experiments \cite{cirigliano_non-standard_2013, biggio_general_2009, severijns_tests_2006, bakhti_sensitivities_2017, blennow_non-unitarity_2017,ohlsson_status_2013, dev_neutrino_2019,papoulias_recent_2019,liao_nonstandard_2017-1,de_gouvea_non-standard_2016,liao_nonstandard_2017, biggio_loop_2009}. Among these have been particle decay rates \cite{cirigliano_non-standard_2013, biggio_general_2009, severijns_tests_2006}, neutrino mixing parameters \cite{bakhti_sensitivities_2017, blennow_non-unitarity_2017, dev_neutrino_2019}, neutrino CP-violating phases \cite{bakhti_sensitivities_2017, blennow_non-unitarity_2017, dev_neutrino_2019}, scattering cross-sections \cite{ohlsson_status_2013, dev_neutrino_2019}, and matter-affected oscillation probabilities \cite{ohlsson_status_2013, dev_neutrino_2019}. This body of research has constrained NSI (both charged-current and neutral-current) to be anywhere on the order of $\mathcal{O}(10^{-4})$ to $\mathcal{O}(10^{-1})$, depending on the data utilized for the analysis and on the NSI model chosen. These constraints tend to be more strict with CC NSI.

Our work here is motivated by another potential observable for NSI. We focus on reported perturbations in nuclear decay rates previously unexplored specifically in the NSI framework. Evidence suggestive of BSM physics is included in several reports, and some of the principal studies are presented in Refs.\ \cite{jenkins_evidence_2009, jenkins_additional_2012,fischbach_time-dependent_2009,sturrock_analysis_2014, jenkins_perturbation_2009,fischbach_indications_2018}. Table 1 in Ref.\ \cite{nistor_phenomenology_2014} summarizes the majority of the reported decay rate perturbations. These perturbations exhibit periodic fluctuations in the daily, yearly, and $\sim12$-year ranges. There are additionally reports of ``single-event" decay rate perturbations associated with solar storms \cite{mohsinally_evidence_2016} and a binary neutron star inspiral \cite{fischbach_indications_2018}.

At first glance, one feasible explanation for the decay rate perturbations---especially in their annual periodicity---is the effect of instrumentation sensitivity to the environment. For example, it would make sense that temperature-dependent effects would be manifested in annual perturbations in the detector counting rate. While this could explain some of the decay rate deviations from the expected behavior, Ref.\ \cite{jenkins_analysis_2010} argued that external influences such as temperature could not sufficiently explain the entirety of the fluctuations. However, there are still criticisms of the possibility that something external to instrumentation errors could influence decay rates. For example, Ref.\ \cite{pomme_evidence_2016} argues against the influence of the Sun on decay rates.

Regardless, it is intriguing that most of the decay rate perturbations have been associated with beta-minus decays specifically, as evident in Table 1 of Ref.\ \cite{nistor_phenomenology_2014}. Furthermore, the decay rate perturbations appear at the $\mathcal{O}(10^{-3})$ level for a variety of independent experiments. Finally, these perturbations appear to be correlated with a varying neutrino flux through the decaying sample. One reported connection between periodic fluctuations and a varying neutrino flux was presented in Ref.\ \cite{sturrock_comparative_2016} in which the authors reference both annual neutrino flux variations at the Super-Kamiokande experiment as well as longer (e.g., $12.5$ year$^{-1}$) variations suggestive of periodic internal solar processes (e.g., the rotation of the radiative zone).

With evidence suggestive of decay rate perturbations correlated with varying neutrino fluxes, we ask ourselves the following question: What can the NSI framework say about beta-minus decay rate perturbations? In answering this question, we aim to construct a more complete mathematical framework to model the decay perturbations. Our goal is to construct a framework for which there would be experimental sensitivity to the model parameters. To accomplish this, we will extend our analysis of decay perturbations beyond a dependence solely on local neutrino flux to include parameters such as neutrino energy and flavor. Finally, we will explore what NSI constraints can be determined using current and future data from a nuclear reactor experiments.



\section{Formalism for CC NSI and Beta Decays}

In this study, we are focused on beta-minus decay rate NSI. For this reason, we will concentrate on the area of quark-neutrino dominating charged-current (CC) NSI. To model these NSI, we introduce a relative strength to the quark-neutrino dominating interaction of beta-minus decays. This is achieved with a modification $\mathcal{L}_\text{NSI}^q$ to the Standard Model Lagrangian, given by

\begin{equation}
    \mathcal{L}_\text{SM + NSI}^q = \mathcal{L}_\text{SM}^q + \mathcal{L}_\text{NSI}^q.
\end{equation}

\noindent In general, we expect $\mathcal{L}_\text{NSI}^q$ to be small compared to $\mathcal{L}_\text{SM}^q$. Specifically, we can then express the CC NSI Lagrangian as

\begin{equation}
    \mathcal{L}_\text{NSI}^q =  -\frac{G_F}{\sqrt{2}} \varepsilon_{\alpha \beta}^{qq'P} V_{qq'} [\bar{q} \gamma^\mu (1 - \gamma_5) q'][\bar{l}_\alpha \gamma_\mu (1-\gamma_5) \nu_\beta] + \text{h.c.}
\end{equation}

In this model, there is a relative CC NSI strength $\varepsilon_{\alpha \beta}^{qq'P} \in \mathbb{C}$ to an electroweak process involving quark flavors $q,q'$ and lepton generations $\alpha, \beta$. The two quark flavors of interest in beta-minus decay are up-type ($q=u$) and down-type ($q'=d$), respectively generated and annihilated. This leaves $V_{qq'}=V_{ud}$ as one element of the Cabibbo-Kobayashi-Maskawa (CKM) quark-mixing matrix. Finally, we are interested in the generation of an electron $\alpha=e$ and the introduction of an electron anti-neutrino $\beta=e$. 

It is important to distinguish between the expression for vector ($\varepsilon_{\alpha \beta}^{udR} + \varepsilon_{\alpha \beta}^{udL})$ and axial-vector ($\varepsilon_{\alpha \beta}^{udR} - \varepsilon_{\alpha \beta}^{udL}$) structures for $P = L, R$. For the beta-minus process here, we will make use of the vector structure

\begin{equation}
    \varepsilon_{\alpha \beta}^{udV} = \varepsilon_{\alpha \beta}^{udR} + \varepsilon_{\alpha \beta}^{udL}. 
\end{equation}

\noindent This all further specifies the new interaction Lagrangian to be

\begin{equation}
    \mathcal{L}_\text{NSI}^q =  -\frac{G_F}{\sqrt{2}} \varepsilon_{ee}^{udV} V_{ud} [\bar{u} \gamma^\mu (1 - \gamma_5) d][\bar{e} \gamma_\mu (1-\gamma_5) \nu_e] + \text{h.c.}
\end{equation}

From this framework, we know that the beta decay rate $\Gamma$ is related to $G_F$ and $V_{ud}$ through

\begin{equation}
    \Gamma \propto G_F^2 |V_{ud}|^2.
\end{equation}

\noindent If we were to consider only the coherent contributions of the NSI to $\Gamma$, we could expect a relationship that is approximately

\begin{equation}
    \delta \Gamma \propto \text{Re}(\varepsilon_{ee}^{udV}),
\end{equation}

\noindent for the difference $\delta \Gamma = \Gamma - \Gamma_0$ of the perturbed rate $\Gamma$ and nominal rate $\Gamma_0$. However, we wish to be as general as possible. For this reason, we will additionally include the incoherent sum of each NSI strength $|\varepsilon_{e\alpha}^{udV}|^2$ for each flavor $\alpha$ to obtain

\begin{equation}\label{eq:gamma_depend}
    \delta \Gamma \propto 2\text{Re}(\varepsilon_{ee}^{udV}) + \sum_\alpha |\varepsilon_{e\alpha}^{udV}|^2.
\end{equation}

\noindent This is the parameterization used in Ref.\ \cite{biggio_general_2009} as they consider NSI bounds using CKM unitarity and experimental determinations of $G_F$.

\section{NSI Constraints from Beta Decay Rates}   

We will take two approaches to constrain the NSI. First, we will take into account only the uncertainty $\delta \Gamma$ in beta-minus decay rates. This is accomplished with the following form:

\begin{equation}
    \Gamma_\text{obs} = \Gamma \left(1 + 2\text{Re}(\varepsilon_{ee}^{udV}) + \sum_\alpha |\varepsilon_{e\alpha}^{udV}|^2\right).
\end{equation}

\noindent Here, we can divide the observed decay rate $\Gamma_\text{obs}$ by the predicted rate $\Gamma$. This measured quantity will have a relative uncertainty $\delta \Gamma / \Gamma$ that we will assign based on decay rate perturbation data. Using data from Ref.\ \cite{jenkins_evidence_2009}, we determine that an appropriate relative uncertainty is $\delta \Gamma / \Gamma = 10^{-3}$. 

To constrain $\varepsilon_{ee}^{udV}$ and the other $\varepsilon_{e\alpha}^{udV}$, it is standard practice to only allow one nonzero $\varepsilon$ at a time in the analysis. This is done to avoid cancellations from the sum of negative and positive values that can come from $\text{Re}(\varepsilon_{ee}^{udV})$ and the $|\varepsilon_{e\alpha}^{udV}|^2$. Using a Markov chain Monte Carlo (MCMC) simulation \cite{gelman_bayesian_2014}, we apply the Metropolis-Hastings algorithm for skewed jumping rules to constrain the $|\varepsilon_{e\alpha}^{udV}|$ to be non-negative. Then, the following bounds (95\% confidence interval) can be placed:

\begin{equation}
    \left| \text{Re}(\varepsilon_{ee}^{udV})\right| \leq 8 \times10^{-4},
\end{equation}

\begin{equation}
    \left| \varepsilon_{e\alpha}^{udV} \right| \leq 0.03.
\end{equation}

In the second approach, we will still consider the uncertainty in beta-minus decay rates as from above, but we will also include uncertainties in the Fermi constant and the CKM matrix element $|V_{ud}|$. This is done to determine how sensitive the NSI constraints are to all possible contributions to their strength. We then write down

\begin{equation}
    G_F^2 |V_{ud}|^2 = \Tilde{G_F}^2 \Tilde{|V_{ud}|}^2 \left(1 + 2\text{Re}(\varepsilon_{ee}^{udV}) + \sum_\alpha |\varepsilon_{e\alpha}^{udV}|^2\right).
\end{equation}

\noindent Here, we will follow standard error propagation rules for the following quantities: $\Tilde{G_F} = G_F \pm \delta G_F$, $\Tilde{|V_{ud}|} = |V_{ud}| \pm \delta |V_{ud}|$, and $G_F^2 |V_{ud}|^2 = G_F^2 |V_{ud}|^2(1 \pm \delta \Gamma/\Gamma$). Once more, we will only allow one nonzero $\varepsilon$ at a time in the analysis. We still have $\delta \Gamma / \Gamma = 10^{-3}$ from  Ref.\ \cite{jenkins_evidence_2009}. Additionally, we consider the best-fit value for $|V_{ud}| = 0.97417 \pm 0.00021$ \cite{patrignani_et_al_particle_data_group_review_2016} and the best-fit value for $G_F = (1.166378 \pm 0.0006)\times 10^{-5} \text{ GeV}^{-2}$ \cite{patrignani_et_al_particle_data_group_review_2016}. As before, we utilize MCMC to determine the following bounds (95\% confidence):

\begin{equation}
    \left| \text{Re}(\varepsilon_{ee}^{udV})\right| \leq 0.001,
\end{equation}

\begin{equation}
    \left| \varepsilon_{e\alpha}^{udV} \right| \leq 0.04.
\end{equation}


\noindent These two methods produce comparable bounds, and they are similar to those of $\mathcal{O}(10^{-4})$ to $\mathcal{O}(10^{-2})$ for a generic quark-neutrino dominating CC NSI constrained both by the Fermi constant and by assuming CKM unitarity as in Ref.\ \cite{biggio_general_2009}.

\section{Model for Beta-Minus Decay NSI}

Next, we aim to further parameterize the CC NSI. Previous studies have explored a perturbed decay rate of the form $\Gamma = \Gamma_0(1 + \Delta)$ in beta decays (e.g., in Section 4 in Ref.\ \cite{barnes_upper_2019}). This modified decay rate has been generally defined as a function of neutrino flux through a sample, i.e., $\Delta = \Delta(F_\nu)$. However---as we alluded to before---there has not been in-depth consideration of dependence on other parameters. Namely, this includes neutrino flavor, type (antimatter or matter), energy, and source. 

\subsection{New Parameters for Beta-Minus Decay NSI}

According to evidence suggestive of time-varying beta-minus decay rates correlated with a varying local neutrino flux (e.g., Refs.\ \cite{jenkins_evidence_2009, jenkins_additional_2012,fischbach_time-dependent_2009,sturrock_analysis_2014, jenkins_perturbation_2009,fischbach_indications_2018}), we draw attention to the following factors for neutrinos involved in the CC NSI : \\

\textsc{flux through decaying sample}---comparable to solar neutrino flux, \\

\textsc{energy}---relatively low $\sim0.1$--$1$ MeV, \\

\textsc{type}---neutrinos as opposed to anti-neutrinos, \\

\textsc{source}---created within a volume of radius much larger than the oscillation length 

(Sun, cataclysmic events), \\



\textsc{decay process(es) affected}---beta-minus decay processes in isotopes summarized 

in Table 1 of Ref.\ \cite{nistor_phenomenology_2014}. \\

Furthermore, there is insufficient evidence for both time-varying decay rates correlated with varying reactor anti-neutrino flux \cite{barnes_upper_2019} and for a ``self-induced decay" (SID) effect in Au-198 \cite{lindstrom_absence_2011, lindstrom_study_2010}. Briefly, the SID effect would describe perturbations in a radioactive sample's decay rate induced (or suppressed) by its own neutrino flux. These reports solidify constraints on neutrino energy, type, and source. Therefore, several sources of data converge upon a decently well-defined, consistent set of factors. Acknowledging these factors, we can move forward to a more specific mathematical framework for the CC NSI---one that includes more relevant parameters.

\subsection{Further Parameterization of Beta-Minus Decay NSI}

We will take into account neutrino flux as before. Additionally, we will include separate neutrino flavor contributions as well as neutrino energy and type (neutrino vs. anti-neutrino). To begin, we consider an electron neutrino's contribution $\varepsilon_e$ to the NSI effect $\varepsilon_{ee}^{udV}$ we constrained in Section 3. We will define $f_e = f_e(E_e, F_e) \in \mathbb{R}$ as the unknown function that relates the energy $E_e$ and flux $F_e$ to the strength $\varepsilon_e \in \mathbb{C}$ of the NSI to obtain

\begin{equation}
    \varepsilon_e(E_e, F_e) \propto f_e(E_e, F_e).
\end{equation}

To include other flavor contributions to the NSI, we will introduce a weighted sum. This takes us from only the electron neutrino contribution $\varepsilon_e$ to the full NSI effect $\varepsilon$ from all flavors: 

\begin{equation}
     \varepsilon_e \rightarrow \varepsilon(E_e, E_\mu, E_\tau, F_e, F_\mu, F_\tau) \propto \sum_\alpha \beta_\alpha f_\alpha(E_\alpha, F_\alpha)
\end{equation}

\noindent with weights $\beta_\alpha \in \mathbb{C}$. Often, the $E_\alpha$ are all the same value. For convenience, then, we will collapse the notation to $E = E_e = E_\mu = E_\tau$. 

Next, the proportionality factor we will choose is $\varepsilon_{ee}^{udV}$ (i.e., what we constrained in Section 3). This gives us

\begin{equation} \label{eq:sum_def}
    \varepsilon = \varepsilon_{ee}^{udV} \sum_\alpha \beta_\alpha f_\alpha(E, F_\alpha).
\end{equation}

\noindent We acknowledge that we could have chosen different combinations of $\beta_\alpha$ and $f_\alpha$ defined to be in $\mathbb{R}$ or $\mathbb{C}$. However, the parameterization we have chosen should not alter our analysis significantly.

We would like to normalize our $\beta_\alpha$, $f_\alpha$ in some way. For the $f_\alpha$, we will choose $f_\alpha = 1$ around the maximum solar flux $F_\text{solar}$ for neutrinos of flavor $\alpha$ on Earth and for $E$ comparable to the energy $E_\text{solar}$ of solar neutrinos. In this scenario,

\begin{equation}
    \varepsilon(E_\text{solar}, F_\text{solar}) = \varepsilon_{ee}^{udV} \sum_\alpha \beta_\alpha.
\end{equation}

Finally, we normalize the $\beta_\alpha$. In Section 3, we used data with $E = E_\text{solar}$, $F \sim F_\text{solar}$ to arrive at constraints for $\varepsilon_{ee}^{udV}$. This means that we had $\varepsilon(E_\text{solar}, F_\text{solar}) = \varepsilon_{ee}^{udV}$. This requires $\sum_\alpha \beta_\alpha = 1$ so that

\begin{equation}
    \text{Re} \left[\varepsilon(E_\text{solar}, F_\text{solar})\right] = \text{Re}(\varepsilon_{ee}^{udV})
\end{equation}

\noindent and

\begin{equation}
    |\varepsilon(E_\text{solar}, F_\text{solar})| = |\varepsilon_{ee}^{udV}|.
\end{equation}

For both neutrino and anti-neutrino effects, we would have the generalized beta-minus decay NSI contributions

\begin{equation}
    \varepsilon = \varepsilon_{ee}^{udV} \sum_\alpha \big( \beta_\alpha f_\alpha(E, F_\alpha) + \gamma_\alpha g_\alpha (\bar{E}, \bar{F}_\alpha)\big),
\end{equation}


\noindent where the $\gamma_\alpha, g_\alpha, \bar{E}, \bar{F}_\alpha$ are the corresponding anti-neutrino parameters and properties relevant to the CC NSI. From existing experimental data, the evidence suggests that the $\gamma_\alpha g_\alpha \ll \beta_\alpha f_\alpha$ (e.g., the strictly constrained signal in Ref.\ \cite{barnes_upper_2019} with reactor anti-neutrinos). Therefore---in what follows---we will only analyze the neutrino contributions to the NSI and not those of anti-neutrinos. This means that our model takes the form of Eq.\ (\ref{eq:sum_def}).

\subsection{Discussion of Model Choice for Beta-Minus Decay Rate NSI}

At this point, a potential problem might be apparent. Indeed, we could have generated a separate parameterization for the other $\varepsilon_{e\alpha}^{udV}$ since we only walked through this process with $\varepsilon_{ee}^{udV}$. This could have been accomplished by adding 3 more $\beta$ parameters and $f$ functions for $\varepsilon_{e\mu}^{udV}$ and another 3 for $\varepsilon_{e\tau}^{udV}$. If we wanted to pursue the most general parameterization, this would be one route to take. However, we will continue under the assumption that only $\varepsilon_{ee}^{udV}$ contributes to the NSI effect with our parameterization.

With this approach, we will allow all incoherent contributions to $\Gamma$ from the $|\varepsilon_{e\alpha}^{udV}|^2$ to exist. Then, the one new stipulation is that $\varepsilon_{ee}^{udV}$ is a function of the new parameters assigned in this section. If we were incorrect to give the $\varepsilon_{ee}^{udV}$ this structure, data analysis would reveal that, for example, the $f_\alpha$ do not change with varied $E$ and the $F_\alpha$. This would imply that the NSI would not change with $E$ and the $F_\alpha$. In the next section, we will explore a concrete methodology for an experiment that could provide the data needed to probe the new NSI parameter space.

\section{Sensitivity from Reactor Neutrino Experiments}

One way in which we could expect to constrain our model parameters would be with reactor experiments at a short range. In this scenario, the electron neutrino survival probability $P_{ee}$ is arbitrarily close to 1, i.e., $P_{e\mu} \approx P_{e\tau} \approx 0$ such that $F_\mu, F_\tau \ll F_e$. This would lead to a measurement of $\varepsilon \rightarrow \varepsilon_e$ in which the effects of $\nu_\mu$ and $\nu_\tau$ are negligible. 

To illustrate this, we will assume that the $\nu_e$ energy and flux contribute independently to $f_e$ around the solar neutrino flux and energy. This will allow us to break down the function $f_e$ into manageable pieces for the energy contribution we will call $X_e$ and the flux contribution we will call $Y_e$:

\begin{equation}
    f_e(E, F_e) \rightarrow X_e(E) Y_e(F_e) \approx X_{e,0} Y_e(F_e) \text{ for } E \sim E_\text{solar},
\end{equation}

\noindent where we have the nominal energy contribution $X_{e,0}$. This is convenient because we required in Section 4 that $f_e = 1$ when $E = E_\text{solar}$, $F \sim F_\text{solar}$. In other words, we should have $X_{e,0} = 1$ from this normalization. With all of this, we can then rewrite Eq.\ (\ref{eq:sum_def}) as

\begin{equation}
    \varepsilon \rightarrow \varepsilon_{e} = \varepsilon_{ee}^{udV} \beta_e Y_e.
\end{equation}

\noindent With this expression, it will be possible to explicitly see the effects of $\beta_e$ and $Y_e$ on the beta decay rate within Eq.\ (\ref{eq:gamma_depend}):


\begin{equation}
    \delta \Gamma \propto 2Y_e\text{Re}(\varepsilon_{ee}^{udV} \beta_e) + Y_e^2|\varepsilon_{ee}^{udV} \beta_e|^2 + |\varepsilon_{e\mu}^{udV}|^2 + |\varepsilon_{e\tau}^{udV}|^2.
\end{equation}


Electron neutrino fluxes at the centers of nuclear reactors---for the sake of neutrino physics experiments---can be at least $F_e(L=0) = 10^{14}$ cm$^{-2}$ s$^{-1}$ (e.g., see Ref.\ \cite{texono_collaboration_production_2005}). For $E \approx E_\text{solar}$, the $\nu_e$ survival probability is rather large at short distances, since oscillations are generally relevant only for $L/E > 100$ m MeV$^{-1}$. The required distance to reproduce the solar neutrino flux would be on the order of $L_0 \sim 10$ m, assuming spherically emitted neutrinos, i.e., $F_e(L) = F_e(L=0) / 4 \pi L^2$. For a practical example, the experiment in Ref.\ \cite{barnes_upper_2019} intercepted electron anti-neutrinos around $5$ m from the nuclear reactor at which the anti-neutrino flux was $\sim50$ times the solar neutrino flux.


Since we already know that $X_{e,0} = 1$ from the normalization in Section 4, we also know that $Y_e = 1$ when $F_e$ is at the solar neutrino flux. For small perturbations around $Y_e$, then, we could expect a first-order effect:

\begin{equation}
    Y_e \approx 1 + y_e(F_e - F_{e,0}),
\end{equation}

\noindent which leaves us with

\begin{equation}
    \varepsilon_{e} = \varepsilon_{ee}^{udV} \beta_e \left( 1 + y_e(F_e - F_{e,0}) \right) = \varepsilon_{ee}^{udV} \beta_e ( 1 + y_e \Delta F_e).
\end{equation}

\subsection{Constraining Electron Neutrino CC NSI Contributions}

The experiment will need to reproduce the variation in annual neutrino flux to probe BSM physics from studies such as Ref.\ \cite{jenkins_evidence_2009}. This would be achieved by varying the detector distance up to $6\%$. This percentage comes from the comparison of decay rate perturbations to the Earth-Sun distance \cite{jenkins_evidence_2009}. This corresponds to a change in detector position on the order of $\delta L \sim 1$ m, and the detector displacement can be performed as in Ref.\ \cite{barnes_upper_2019}. At this distance, we can constrain $\beta_e$ and $y_e$ with a ratio of decay rates:


\begin{multline}
    \frac{\Gamma(\Delta F_e = 0)}{\Gamma(\Delta F_e)} = \Big(1 + \text{Re}(\varepsilon_{ee}^{udV} \beta_e) + |\varepsilon_{ee}^{udV} \beta_e|^2 + |\varepsilon_{e\mu}^{udV}|^2 + |\varepsilon_{e\tau}^{udV}|^2\Big) \\
    \times \Big(1 + Y_e \text{Re}(\varepsilon_{ee}^{udV} \beta_e) + Y_e^2 |\varepsilon_{ee}^{udV} \beta_e|^2 + |\varepsilon_{e\mu}^{udV}|^2 + |\varepsilon_{e\tau}^{udV}|^2\Big)^{-1}.
\end{multline}

\noindent Once more, we take only one nonzero $\varepsilon$ at a time for the analysis. A convenient route to take with the constraints we already have on the $|\varepsilon_{e\alpha}^{udV}|$ is to consider nonzero $|\varepsilon_{ee}^{udV} \beta_e|$:

\begin{equation}\label{eq:gamma_compare}
\frac{\Gamma(\Delta F_e = 0)}{\Gamma(\Delta F_e)} \rightarrow \frac{1 + |\varepsilon_{ee}^{udV} \beta_e|^2}{1 + Y_e^2 |\varepsilon_{ee}^{udV} \beta_e|^2} = \frac{1 + |\varepsilon_{ee}^{udV}|^2 |\beta_e|^2}{1 + (1 + y_e \Delta F_e)^2 |\varepsilon_{ee}^{udV}|^2 | \beta_e|^2}.
\end{equation}

As we mentioned in Section 4, finding no change to the $f_\alpha$ as a function of $E$ and $F_\alpha$ would be evidence against parameterizing $\varepsilon_{ee}^{udV}$ in the way we have done. Here, then, we need to look specifically at $f_e$, which changes with the parameter $y_e$. The absence of evidence suggestive of a nonzero $y_e$ parameter would disfavor our $\varepsilon_{ee}^{udV}$ parameterization.

We will take two approaches to constrain $\beta_e$ and $y_e$. The first is to determine what sensitivity current experimental technology would have if decay rate data were constant with $\mathcal{O}(10^{-3})$ statistical fluctuations. This is similar to assuming that the time-varying decay rate data measured in Refs.\ \cite{jenkins_evidence_2009, jenkins_additional_2012,fischbach_time-dependent_2009,sturrock_analysis_2014, jenkins_perturbation_2009,fischbach_indications_2018} are attributable entirely to external factors like temperature (however---as we pointed out---it was demonstrated that these would not likely be the only cause of the time-varying fluctuations \cite{jenkins_analysis_2010}). 

Running an MCMC simulation for $\Gamma(\Delta F_e=0)/\Gamma(\Delta F_0) = 1.000\pm0.001$, we find

\begin{equation}\label{eq:beta_same}
    |\beta_e| \leq 0.4 \text{ (95\%)},
\end{equation}


\noindent and no significant sensitivity to $y_e$. These bounds mean that current experimental technology involving only $\nu_e$ could favor a lower mixing constant for $\nu_e$ NSI contributions while not favoring a particular value for $y_e$.


The second approach is to determine what sensitivity current experiments would have if the fluctuations in Refs.\ \cite{jenkins_evidence_2009, jenkins_additional_2012,fischbach_time-dependent_2009,sturrock_analysis_2014, jenkins_perturbation_2009,fischbach_indications_2018} were not at all attributable to instrumental or environmental effects. In this scenario, another MCMC simulation shows the following for $\Gamma(\Delta F_e=0)/\Gamma(\Delta F_0) = 1.004\pm0.001$:

\begin{equation}\label{eq:beta_diff}
    |\beta_e| \in [0.5, 1.6] \text{ (95\%)}, 
\end{equation}

\begin{equation}
    y_e \in \left[-0.97\frac{F_{e,0}}{\Delta F_e}, -0.07\frac{F_{e,0}}{\Delta F_e}\right] \text{ (95\%)},
\end{equation}


\noindent for $\Delta F_e / F_{e,0} = 0.06$. These bounds would favor heavy NSI contributions from $\nu_e$ as well as specific, nonzero $y_e$ values if the data from Ref.\ \cite{jenkins_evidence_2009} were beta decay rate effects from only $\nu_e$. For $\beta_e$ and $y_e$, then, we find that reactor neutrino experiments could be sensitive to our NSI parameters.

\subsection{Constraining Muon and Tau Neutrino Contributions}

We can additionally constrain $\beta_\mu$, $\beta_\tau$ using the requirement $\sum_\alpha \beta_\alpha = 1$ from before:

\begin{equation}
    |\beta_e| = |1 - \beta_\mu - \beta_\tau| \\ = \sqrt{\big(1 - \text{Re}(\beta_\mu + \beta_\tau)\big)^2 + \text{Im}(\beta_\mu + \beta_\tau)^2}.
\end{equation}

\noindent With the estimated posterior for $|\beta_e|$, we will allow for free parameters $\text{Re}(\beta_\mu + \beta_\tau)$ and $\text{Im}(\beta_\mu + \beta_\tau)$. The following constraints are determined using an MCMC simulation with the bounds from Eq.\ (\ref{eq:beta_same}) in which $\Gamma(\Delta F_e=0)/\Gamma(\Delta F_0) = 1.000\pm0.001$:

\begin{equation}
    \text{Re}(\beta_\mu + \beta_\tau) \in [0.6, 1.5] \text{ (95\%)},
\end{equation}

\begin{equation}
    \text{Im}(\beta_\mu + \beta_\tau) \in [-0.4, 0.4] \text{ (95\%)}.
\end{equation}

\noindent Next---for the bounds from Eq.\ (\ref{eq:beta_diff}) in which $\Gamma(\Delta F_e=0)/\Gamma(\Delta F_0) = 1.004\pm0.001$---we use MCMC once more to obtain the following constraint:

\begin{equation}
    \big(\text{Re}(\beta_\mu + \beta_\tau) -1.0 \big)^2 + \text{Im}(\beta_\mu + \beta_\tau)^2 \in [0.25, 1.0] \text{ (95\%)}.
\end{equation}




With the statistical approaches we introduced in this section, we have arrived at constraints for the $\beta_\alpha$ and for $y_e$. Therefore, the model we have proposed could be reasonably constrained by current or near-future nuclear reactor technology.

\section{Summary and Conclusions}

In summary, we used the CC NSI framework to model time-varying beta decay rates by assuming a neutrino-quark dominating effect. First, we related the coherent and incoherent contributions of the NSI $\varepsilon_{e\alpha}^{udV}$ to the beta decay rate with $\Gamma_\text{obs} \propto 1 + 2\text{Re}(\varepsilon_{ee}^{udV}) + \sum_\alpha |\varepsilon_{e\alpha}^{udV}|^2$. Using data on beta-minus decays, the Fermi coupling constant $G_F$, and the CKM matrix element $V_{ud}$, we determined that the $\varepsilon$ were constrained at $\mathcal{O}(10^{-4})$ to $\mathcal{O}(10^{-2})$, which is consistent with bounds placed in existing literature. Next, we constructed a model based on potential parameters governing the new CC NSI. This model considered the NSI strength $\varepsilon$ to be a function of neutrino flavor, energy, and local flux. We allowed the energy and flux of each flavor $\alpha$ to contribute nominally to the NSI with strength $f_\alpha$, and then these nominal strengths were each weighted by a factor $\beta_\alpha \in \mathbb{C}$ to obtain $\varepsilon = \varepsilon_{ee}^{udV} \sum_\alpha \beta_\alpha f_\alpha$. We determined that current technology for nuclear reactors could produce neutrino experiments sensitive to the $\beta_\alpha$ and to $y_e$. These experiments would study electron neutrino production at short distances such that other flavor contributions would be negligible to the NSI.

From this work, we found that it is possible to parameterize CC NSI in order to model beta-minus decay rate perturbations as functions of several neutrino properties. These parameters could be constrained by existing data and potentially by new experiments using current reactor technology. However, there remain several open questions. For example, it would be instructive for future studies to consider whether the origin or oscillatory nature of neutrinos had any additional effect on the NSI framework presented here. Since solar neutrino oscillations are much different from terrestrial oscillations produced in experiments, the solar neutrino flux could engage in different NSI that we did not take into account. Regardless, the feasibility for NSI frameworks to complete our understanding of neutrino physics is a promising area for further research.

\acknowledgments

We would like to thank Ephraim Fischbach and Dennis Krause for insightful discussions about neutrino physics and about research concerning beta decay rate perturbations.

\end{document}